\documentclass[times,dvipdfmx]{akari2017} 

\shorttitle{AKARI colors of O-rich/C-rich Miras}
\shortauthors{N. Matsunaga}

\begin{document}

\title{AKARI color useful for classifying chemical types of Miras} 



\correspondingauthor{Noriyuki Matsunaga}
\email{matsunaga@astron.s.u-tokyo.ac.jp}

\author{Noriyuki Matsunaga} 
\affiliation{Department of Astronomy, The University of Tokyo, 7-3-1 Hongo, Bunkyo-ku, Tokyo 113-0033, Japan} 



\begin{abstract}
The AKARI/IRC color combining the $S9W$ and $L18W$ bands is useful for
distinguishing between oxygen-rich and carbon-rich circumstellar dust.
Ishihara and collaborators found in 2011 that this color,
used together with the near-IR color $J-K_{\rm s}$,
can be used to classify two groups of dust-enshrouded stars
with different chemistry. They investigated the distributions
of such dusty AGB stars in the Galaxy and found that
those with oxygen-rich dust are more centrally concentrated.
While this is consistent with previous studies, the map in Ishihara et~al.\  
shows that carbon-rich stars are also present in the Galactic bulge
for which almost no carbon-rich stars were reported before.
Here we focus on Mira variables whose distances can be well constrained
based on the period-luminosity relation. Among some candidates of
carbon-rich Miras selected by the AKARI color, we confirmed at least
four carbon-rich Miras within the bulge with their optical spectra.
This gives a new insight into the complicated nature of
stellar populations in the bulge. 
\end{abstract}



\setcounter{page}{1}



\section{Introduction} 

Miras are pulsating stars with long period ($P \geq 100$~d)
and large amplitude ($\geq 2.5$~mag in $V$).
They appear at the last stage of the Asymptotic Giant Branch (AGB),
and their period-luminosity relation serves as a good distance scale
\citep[][and references therein]{Whitelock-2008,Whitelock-2013}.
During their evolution, thermal pulses can dredge up carbon produced
in the interior to the surface, and some of them (and also
some non-Mira AGB stars) get more carbon than oxygen on the surface.
Such stars show very different molecules from others with more oxygen;
for example, C$_2$ for carbon-rich (C-rich) stars, and
H$_2$O and TiO for oxygen-rich (O-rich) stars, in addition to
CO molecules which are present in both groups.  
C-rich AGB stars with this {\it canonical} evolutionary path
are aged at around 0.5--5~Gyr and thus represent intermediate-age
stellar population \citep{Mouhcine-2003,Marigo-2008}.
There are other evolutionary paths for producing C-rich stars
and their ages may be largely different
\citep[see e.g.\ ][]{McClure-1990,Green-1994,Izzard-2007}.

Our focus is C-rich Miras in the Galactic bulge.
Such objects would give us an insight into stellar population
in the bulge and its evolution, although very few C-rich objects
are known in this inner region of the Galaxy.
A small number of C-rich giants were found in early 1990s
\citep{Azzopardi-1991,Tyson-1991}, but their nature is still unclear
\citep{Whitelock-1993,Ng-1997}. They are fainter than
expected for Miras or the late-stage AGBs in the bulge,
and evolutionary paths through binary interaction have been suggested.
\citet{Miszalski-2013} recently discovered a C-rich symbiotic Mira
towards the bulge, although its membership to the bulge was questioned
(also see below). The vast majority of Miras and late-stage AGBs
in the bulge are considered to be O-rich. In order to find C-rich Miras
among a large number of O-rich ones, it is necessary to identify
good candidates of C-rich objects in an efficient way.

\begin{figure}[!ht]
\begin{center}
         \includegraphics[bb=0 0 451 391]{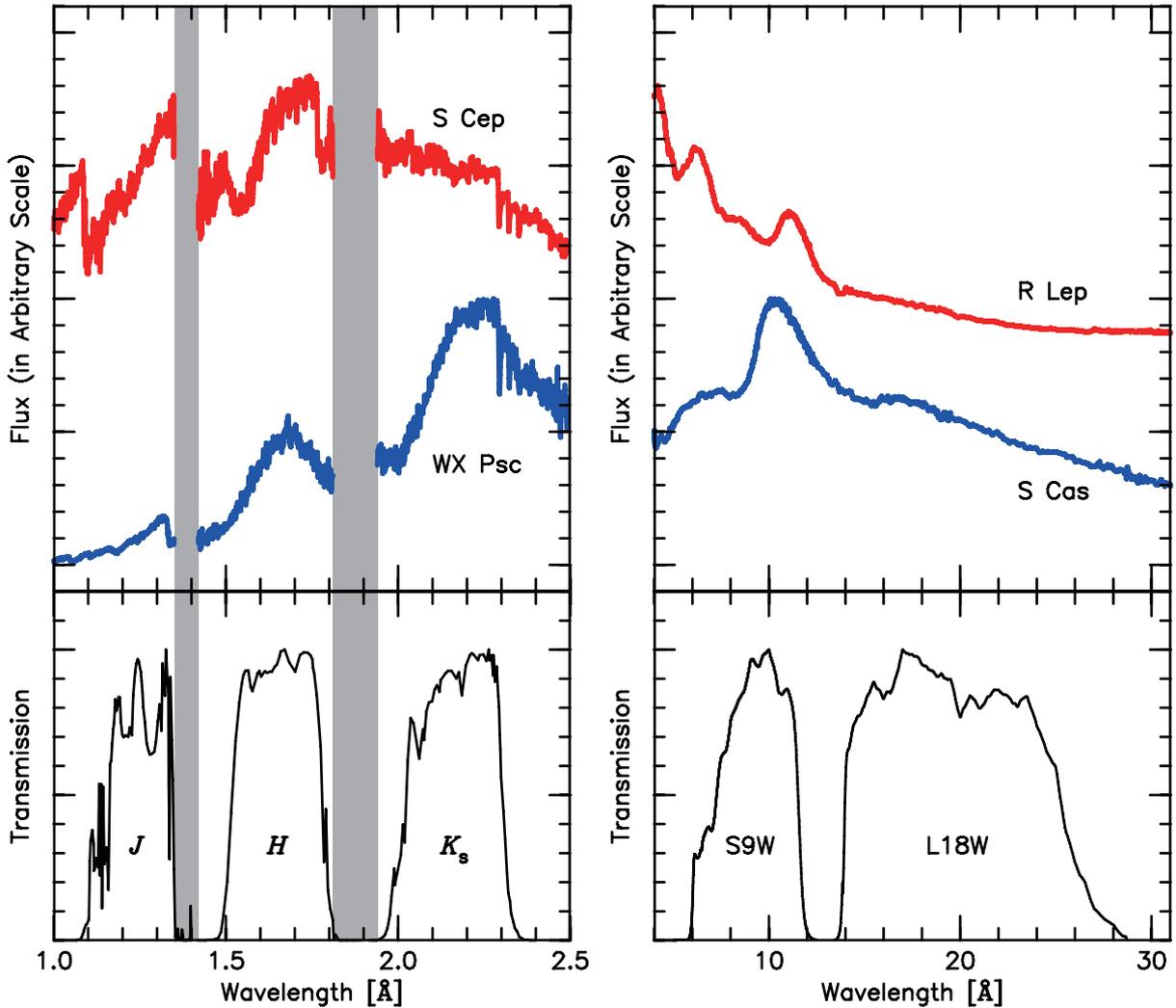}
        \caption{Near-IR spectra from \citet{Lancon-2000} and mid-IR spectra from \citet{Sloan-2003} are presented for C-rich (R~Lep with $P=427$~d and S~Cep with $P=487$~d) and O-rich Miras (WX~Psc with $P=660$~d and S~Cas with $P=608$~d). Also presented are filter transmission curves for 2MASS $JHK_{\rm s}$ bands \citep{Carpenter-2001} and AKARI $S9W$ and $L18W$ bands \citep{Onaka-2007}. Vertical strips in the left panels indicate the heavy telluric absorption between the $JHK_{\rm s}$ bands.
        }\label{fig1}
\end{center}
\end{figure}

On the $(J-K_{\rm s})$-$([9]-[18])$ diagram, C-rich and O-rich stars
are separated into different color locations \citep{Ishihara-2011}. 
In this paper, $[9]$ and $[18]$ stand for magnitudes in
$S9W$ and $L18W$ bands, respectively, taken from
the AKARI/IRC mid-IR all-sky survey catalog \citep{Ishihara-2010},
while the near-IR colors, $J-K_{\rm s}$ were taken from the 2MASS catalog
\citep{Skrutskie-2006}.
The separation is mainly attributed to different color trend
in the mid IR due to different chemical composition
of their circumstellar dust shells.
According to \citet{Ita-2010}, dusty C-rich stars often show
the SiC emission band at 11.3~{$\mu$m} within the [9] band
in addition to a continuum excess due to amorphous carbon dust, 
while dusty O-rich stars have the silicate bands at 9.8 and 18~{$\mu$m}.
In the near-IR bands, C-rich AGB stars have strong C$_2$ and CN bands
in $J$, whilst there are no correspondingly strong bands in $K_{\rm s}$.
In contrast, O-rich AGB stars have strong H$_2$O absorption
which affects both $J$ and $K_{\rm s}$ bands in addition to
TiO and VO bands seen in $J$. Figure~\ref{fig1} presents
near-IR and mid-IR spectra of C-rich and and O-rich Miras
showing their features well.



\section{Selection and spectroscopic follow-up of C-rich Miras} 

\begin{figure}[!ht]
\begin{center}
        \includegraphics[bb=0 0 389 298]{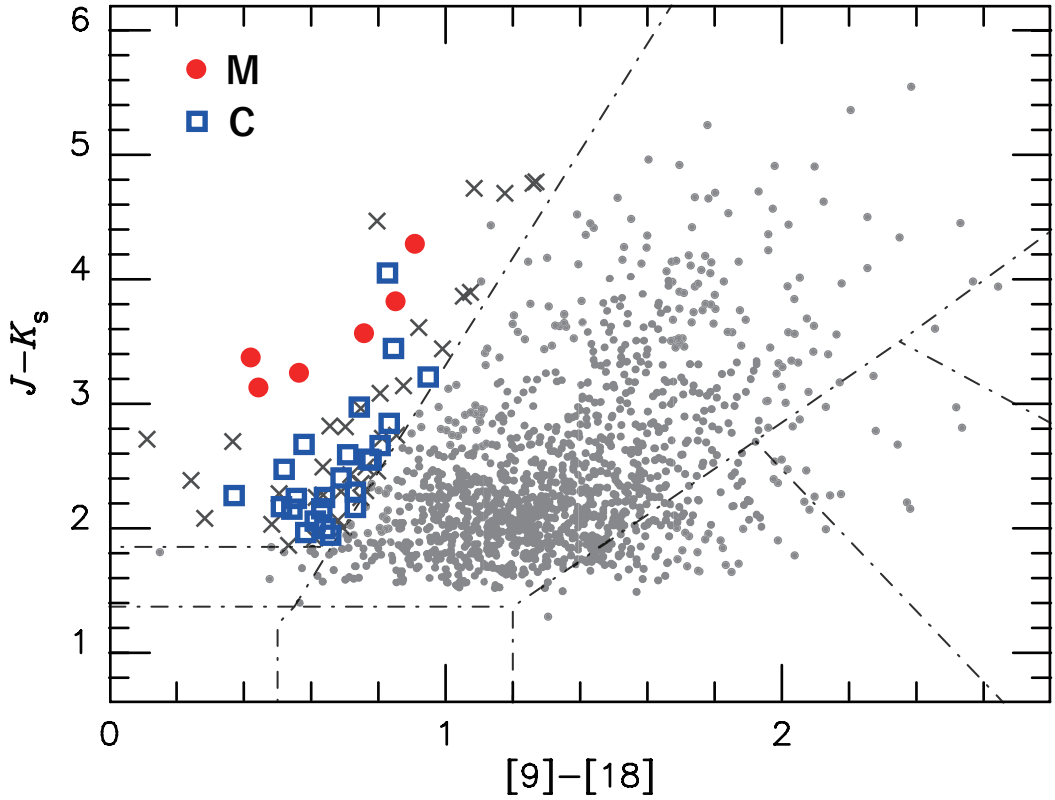}
        \caption{Color-color diagram, $(J-K_{\rm s})$ versus $([9]-18])$, for OGLE-III Miras in the bulge indicated by gray dots. Our targets found to be C-rich and O-rich are indicated by filled circles in red and open squares in blue, respectively, while crosses indicate those not-observed or unclassified. 
        }\label{fig2}
\end{center}
\end{figure}

Based on the $(J-K_{\rm s})$-$([9]-[18])$ diagram (Figure~\ref{fig2}),
we identified 66 candidates of C-rich Miras among more than 6500 Miras
in the OGLE-III catalog of long-period variables in the bulge
\citep{Soszynski-2013}. 33 of them were observed by SpUpNIC 
spectrograph \citep{Crause-2016} attached to the 1.9-m telescope
in South African Astronomical Observatory in 2016 July. 
In addition, we also obtained spectra of 3 Miras in \citet{Catchpole-2016}.
These spectra in the optical range are of low resolution, but
classification between C-rich and O-rich stars is fairly straightforward
with clearly different molecular features of the two groups
(e.g.\  CN and TiO bands for C-rich and O-rich stars, respectively).
Figure~\ref{fig2} plots spectra of eight C-rich Miras we confirmed,
while other 25 Miras were found to be O-rich and poor quality of the spectra
prevented the classification for three. 
We also made near-IR photometric observations of most of these Miras
using the 1.4-m Infrared Survey Facility (IRSF) in
South African Astronomical Observatory, and used the time-series
$JHK_{\rm s}$ magnitudes to check their memberships to the bulge.
Considering their locations on the color-magnitude and the period-wesenheit
diagrams, among the eight C-rich Miras, three
\citep[including the symbiotic Mira in][]{Miszalski-2013} are foreground,
four are very likely within the bulge, while the last one may be background
\citep[see figures in][]{Matsunaga-2017}.
These C-rich Miras in the bulge represents a rare stellar population
in the bulge, although their origin and nature need to be further investigated.
Figure~\ref{fig4} plots their locations in the Galactic coordinate.

\begin{figure}[!ht]
\begin{center}
        \includegraphics[bb=0 0 393 351]{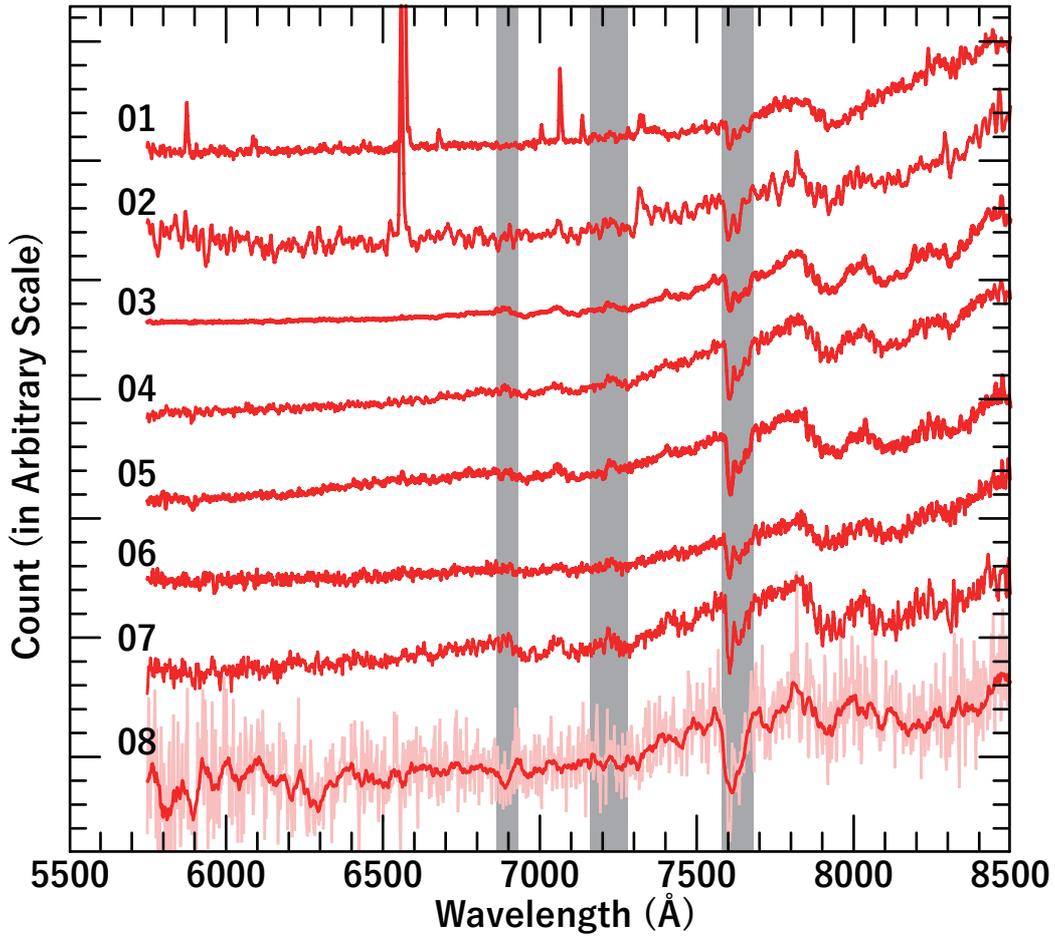}
        \caption{SpUpNIC optical spectra of C-rich Miras we identified. The telluric absorption features (O$_2$ 6867~\AA, H$_2$O 7164~\AA, and O$_2$ 7594~\AA) are shown as shaded vertical bands. The spectrum of No.~08 has a low signal-to-noise ratio as illustrated by the pink curve, but the red curve after smoothing shows the similarity to the other spectra.
        }\label{fig3}
\end{center}
\end{figure}

\begin{figure}[!ht]
\begin{center}
        \includegraphics[bb=0 0 448 344]{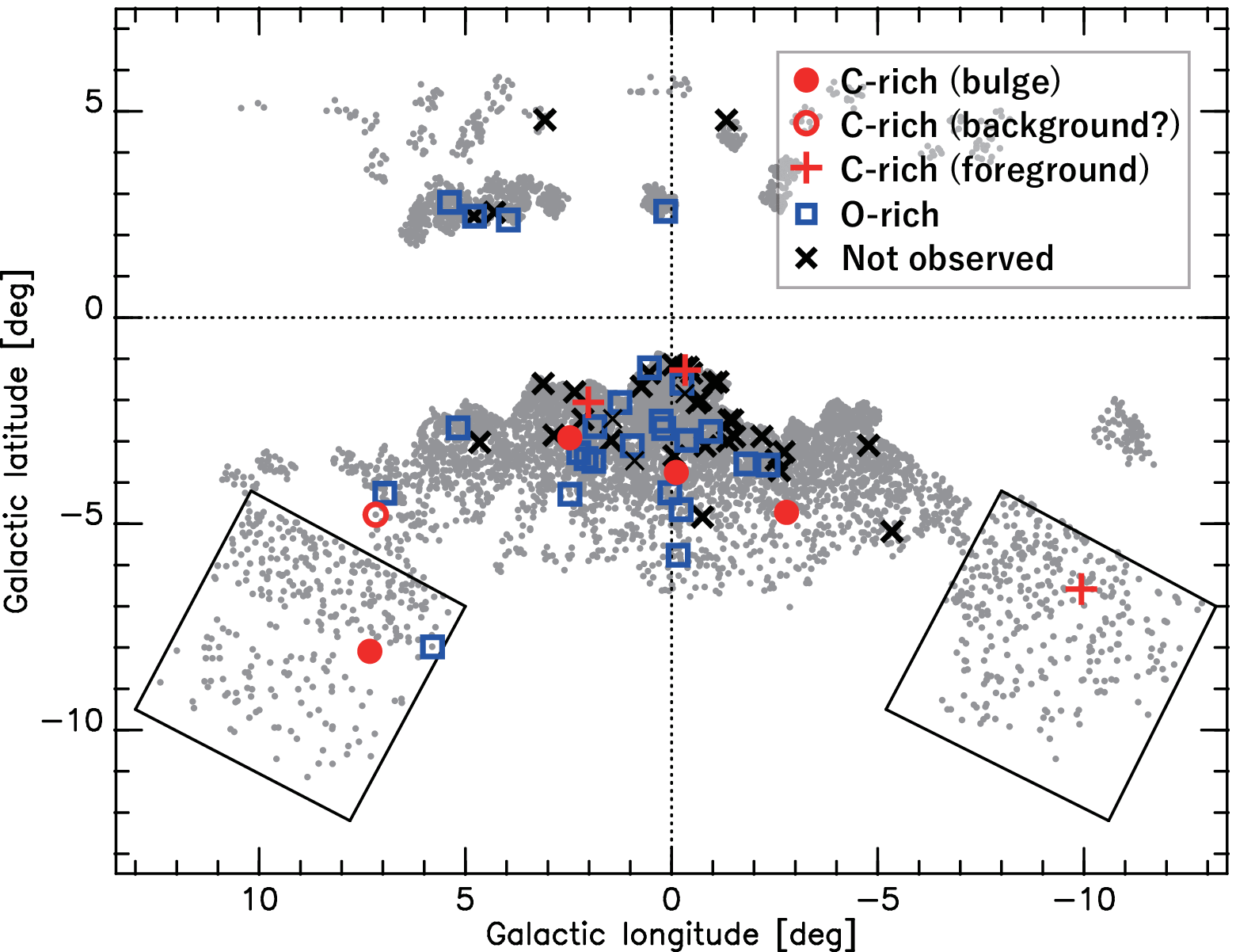}
        \caption{Locations of our targets, eight C-rich Miras (in red as indicated in the legend) but with different distances, O-rich Miras (blue open squares), and those non-observed or classified (black crosses), as well as more than 6500 OGLE-III Miras (grey dots).
        }\label{fig4}
\end{center}
\end{figure}

\section{Concluding remarks}

\begin{figure}[!ht]
\begin{center}
        \includegraphics[bb=0 0 425 270]{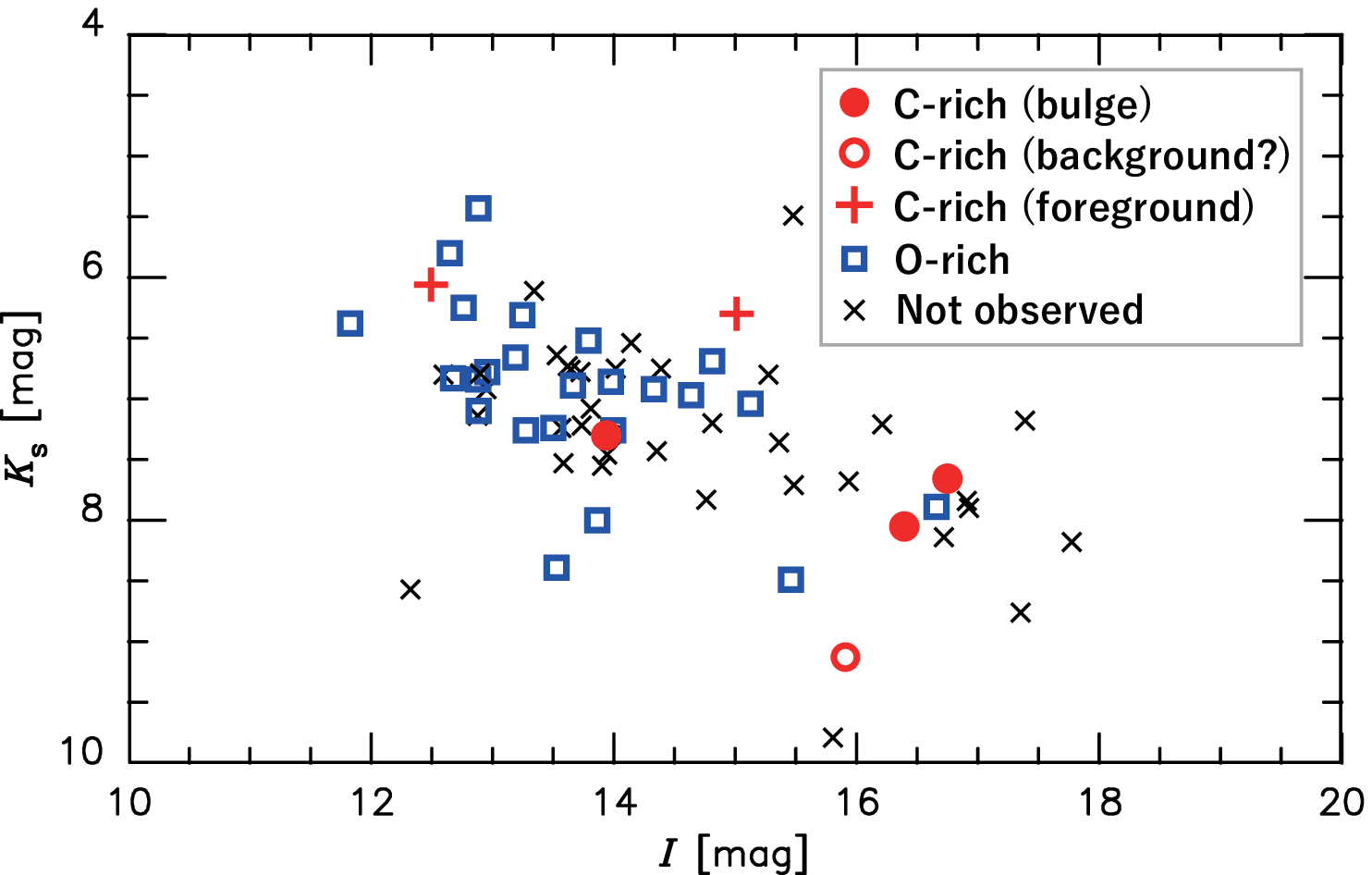}
        \caption{$I$-band and $K_{\rm s}$-band magnitudes are compared for 66 candidates of C-rich Miras (though some are found to be O-rich). 
        }\label{fig5}
\end{center}
\end{figure}

We confirmed that the $(J-K_{\rm s})$-$([9]-[18])$ diagram 
which had been used for general dusty stars \citep{Ishihara-2011}
is also useful for classifying between C-rich and O-rich Miras.
For example, this tool allows one to efficiently identify
rare C-rich Miras in the Galactic bulge where the majority is
O-rich Miras \citep{Matsunaga-2017}. 
The color classification often done by single-epoch photometric catalogs,
however, may be confused by photometric errors and color variation, and
spectroscopic follow-up is important to confirm the diagnosis.
Even low-resolution spectra are useful because of very different
broad molecular bands in the two groups of Miras.
Small-sized telescopes equipped with low-resolution spectrograph
can be useful for such follow-up, but some objects are
highly attenuated by the circumstellar dust (and in some cases by
interstellar dust for objects in the Galactic bulge and disk).
Figure~\ref{fig5} compares $I$-band and $K_{\rm s}$-band magnitudes
of our candidate C-rich Miras in the bulge.
Notably, C-rich Miras tend to be fainter than O-rich ones,
and our spectroscopic observation with SpUpNIC at the 1.9-m SAAO telescope
was limited by the faintness of our targets in $I$. 
Dusty red targets remain relatively bright in $K_{\rm s}$,
indicating the advantage of IR spectrographs.

\subsection*{Acknowledgments}
This contribution is based on a paper by Matsunaga, Menzies, Feast,
Whitelock, Onozato, Barway, \& Aydi (\citeyear{Matsunaga-2017}).
This work was supported as a Joint Research Project under agreement between
the Japan Society for the Promotion of Science (JSPS) and
National Research Foundation (NRF) in South Africa.
NM is grateful to Grant-in-Aid (No.~26287028) from the JSPS.
This research is based on observations with AKARI, a JAXA project with the participation of ESA.





\begin{thebibliography}{}
\bibitem[Azzopardi {et~al.}(1991)]{Azzopardi-1991}
Azzopardi, M., Lequeux, J., Rebeirot, E. et~al. 1991, A\&AS, 88, 265
\bibitem[Carpenter(2001)]{Carpenter-2001}
Carpenter, J.~M. 2001, AJ, 121, 2851
\bibitem[Catchpole {et~al.}(2016)]{Catchpole-2016}
Catchpole, R.~M., Whitelock, P.~A., Feast, M.~W. et~al. 2016, MNRAS, 455, 2216
\bibitem[Crause {et~al.}(2016)]{Crause-2016}
Crause, L. A., Carter, D., Daniels, A. et~al. 2016, Proc.~SPIE, 9908, id.~990827
\bibitem[Green \& Margon(1994)]{Green-1994}
Green, P. J., \& Margon, B. 1994, ApJ, 423, 723
\bibitem[Ishihara {et~al.}(2010)]{Ishihara-2010}
Ishihara, Onaka, T., Kataza, H. et~al. 2010, A\&A, 514, A1
\bibitem[Ishihara {et~al.}(2011)]{Ishihara-2011}
Ishihara, D. , Kaneda, H., Onaka, T. et~al. 2011, A\&A, 534, A79
\bibitem[Ita {et~al.}(2010)]{Ita-2010}
Ita, Y., Matsuura, M., Ishihara, D. et~al. 2010, A\&A, 514, A2
\bibitem[Izzard {et~al.}(2007)]{Izzard-2007}
Izzar,d R.~G., Jeffery, C.~S., \& Lattanzio, J. 2007, A\&A, 470, 661
\bibitem[Lan\c{c}on \& Wood(2000)]{Lancon-2000}
Lan\c{c}on, A., \& Wood, P.~R. 2000, A\&AS, 146, 217
\bibitem[Marigo {et~al.}(2008)]{Marigo-2008}
Marigo, P., Girardi, L., Bressan, A. et~al. 2008,  A\&A, 482, 883
\bibitem[Matsunaga {et~al.}(2017)]{Matsunaga-2017}
Matsunaga, N., Menzies, J.~W., Feast, M.~W. et~al. 2017, MNRAS, 469, 4949
\bibitem[McClure \& Woodsworth(1990)]{McClure-1990}
McClure, R.~D., \& Woodsworth, A.~W. 1990, ApJ, 352, 709
\bibitem[Miszalski {et~al.}(2013)]{Miszalski-2013}
Miszalski, B., Miko{\l}ajewska, J., \& Udalski, A. 2013, MNRAS, 432, 3186
\bibitem[Mouhcine \& Lan\c{c}on(2003)]{Mouhcine-2003}
Mouhcine, M., \& Lan\c{c}on, A. 2003, MNRAS, 338, 572
\bibitem[Ng(1997)]{Ng-1997}
Ng, Y.~K. 1997, A\&A, 328, 211
\bibitem[Onaka {et~al.}(2007)]{Onaka-2007}
Onaka, T., Matsuhara, H., Wada, T. et~al. 2007, PASJ, 59, S401
\bibitem[Skrutskie {et~al.}(2006)]{Skrutskie-2006}
Skrutskie, M.~F., Cutri, R.~M., Stiening, R. et~al. 2006, AJ, 131, 1163
\bibitem[Sloan {et~al.}(2003)]{Sloan-2003}
Sloan, G.~C., Kraemer, K.~E., Price, S.~D. et~al. 2003, ApJS, 147, 379
\bibitem[Soszy\'nski {et~al.}(2013)]{Soszynski-2013}
Soszy\'nski, I., Udalski, A., Szyma\'nski, M.~K. et~al. 2013, AcA, 63, 21
\bibitem[Tyson \& Rich(1991)]{Tyson-1991}
Tyson, N.~D., \& Rich, R.~M. 1991, ApJ, 367, 547
\bibitem[Whitelock(1993)]{Whitelock-1993}
Whitelock, P.~A. 1993, IAUS, 153, 39
\bibitem[Whitelock {et~al.}(2008)]{Whitelock-2008}
Whitelock, P.~A., Feast, M.~W., \& van Leeuwen, F. 2008, MNRAS, 386, 313
\bibitem[Whitelock(2013)]{Whitelock-2013}
Whitelock, P.~A. 2013, IAUS, 289, 209
\end{thebibliography}
\end{document}